\documentclass[pamm,a4paper,fleqn]{w-art}
\usepackage{times,cite,w-thm}
\usepackage[T1]{fontenc}
\usepackage[utf8]{inputenc}
\usepackage{graphicx}
\usepackage{graphics}
\usepackage{epstopdf, epsfig}
\usepackage{amsmath,bm,float}
\newcommand\beq{\begin{equation}}
\newcommand\eeq{\end{equation}}
\newcommand\bea{\begin{eqnarray}}
\newcommand\eea{\end{eqnarray}}
\newcommand\Rey{\mbox{Re}}
\renewcommand{\vec}[1]{\bm{#1}}
\begin{document}
%% \def\leftmark{Session title}
%%
%%    The information for the title page will be placed between
%%    \begin{document} and \maketitle. The order of most entries
%%    is determined by the class file and cannot be changed by
%%    rearranging them. The maketitle command follows after the
%%    abstract.
%%
%%    The following commands will be updated by the publisher:
%%
%%    \renewcommand{\copyrightyear}{2016}
%%    \DOIsuffix{pamm.20161zzzz}
%%    \Volume{16} 
%%    \Year{2016} 
%%    \pagespan{1}{}
%%
%%    The short title is optional:

\TitleLanguage[EN]
\title[Feedback control of ASBL]{Dynamic feedback control through wall suction in shear flows}

\author{\firstname{Moritz} \lastname{Linkmann}%\inst{1,}%
\footnote{Corresponding author: e-mail \ElectronicMail{moritz.linkmann@physik.uni-marburg.de}%, 
     %phone +00\,999\,999\,999,
     %fax +00\,999\,999\,999}
     }} 
%\address[\inst{1}]{\CountryCode[DE]
\address[]{\CountryCode[DE]
Fachbereich Physik, Philipps-University of Marburg,
D-35032 Marburg, Germany
}
\author{\firstname{Bruno} \lastname{Eckhardt}%\inst{1}%
\footnote{Deceased.%, 
     }}
%%
%%    Abstract is required.
\AbstractLanguage[EN]
\begin{abstract}
\emph{This is a dedicatory.} \newline

\noindent
Flow control is of interest in many open and wall-bounded shear flows in order 
to reduce drag or to avoid sudden large fluctuations that may lead to material 
failure. An established means of control is the application of suction through a porous wall. 
Here, we combine suction with a feedback strategy whereby the suction velocity is adjusted
in response to either the kinetic energy or the shear stress at the bottom
wall. The control procedure is then used in an attempt to stabilize invariant
solutions and to carry out direct numerical simulations with a prescribed value
of the friction coefficient.
\end{abstract}
%% maketitle must follow the abstract.
\maketitle                   % Produces the title.

\section{Suction velocity control}
We consider a plane Couette setup with a fluid located in the gap between two
infinitely extended parallel plates at a distance $2d$.
The bottom plate is stationary and the fluid is driven by the motion of the top
plate  with velocity $U_0$ in $x$-direction.  The flow is assumed to
be incompressible and isothermal such that the density can be
regarded as constant.  For large gap widths and a constant suction velocity
$-V_s$ in the wall-normal $y$-direction through the bottom plate, this system
is often used to emulate the asymptotic suction boundary layer (ASBL).  The aim
here is to study the impact of variations in the suction velocity on 
the global properties of the flow.

The stationary laminar solution of the Navier-Stokes equations 
%with no-slip
%boundary conditions at the walls and zero velocity in the spanwise $z$-direction, one
%obtains
is given by
\beq
\label{eq:laminar}
\vec{U} = \begin{pmatrix} \left(1-e^{-yV_s/\nu}\right) \\ -V_s/U_0 \\ 0 \end{pmatrix} \ ,
\eeq
in units of $U_0$, with $\nu$ the kinematic viscosity. The deviations
$\vec{u}$ from the laminar flow then obey the %following evolution 
dimensionless equations
\begin{align}
	\label{eq:momentum}
	\partial_t \vec{u} + \vec{u} \cdot \nabla \vec{u} + \vec{U} \nabla \vec{u} + \vec{u} \nabla \vec{U} + \nabla p & - \frac{1}{\Rey_0}\Delta \vec{u}  = 0 \ , \\
	\label{eq:solenoidal}
        \nabla \cdot \vec{u} & = 0 \ ,
\end{align}
where $p$ is the pressure divided by the constant density $\rho$ and 
$\Rey_0 = U_0 d/\nu$ the Reynolds number based on the velocity of the top plate, the
half-height of the domain and the kinematic viscosity of the fluid.

The feedback control procedure is based on a control variable $R$, here
related to the suction velocity $V_s$ by  $V_s/U_0 = {R}/{R_0^2}$, %}\frac{R}{R_0} \ ,
and an observable $A$, e.g. the
$L_2$-norm $\|\vec{u}\|_2$ or the friction factor $C_f = \tau_w/(\rho U_0^2)$,
where $\tau_w$ is the shear stress at the bottom wall. The dynamics of the control
variable is given by
\begin{align}
	\label{eq:control}
        \dot{R}  & = - \gamma \left(R - R_0\right) - \gamma \mu (A-A_0) \ , 
        %\\
	%\label{eq:coupling}
\end{align}
where $(A_0,R_0)$ are the values of $R$ and $A$ at the desired operating point
%, here
%given by an invariant solution of
%Eqs.~\eqref{eq:momentum}-\eqref{eq:solenoidal}, 
and $\gamma$ and $\mu$ are free parameters. The first parameter 
defines the time scale on which the control is adjusted and,
and second one regulates the sensitivity of the control variable to changes in
the observable. In order to achieve stabilization, the reaction of the control
variable has to be faster than the internal dynamics with which the dynamics
runs away from the operating point, and the amplitude has to be sufficiently
large to switch between stable and unstable regions (details will be given in
\cite{bib:Linkmann2019}).

\section{Numerical aspects and simulation data}
\label{sec:numerics}
Equations ~\eqref{eq:momentum}-\eqref{eq:control} are solved numerically with
an adapted version of the open-source code {\tt channelflow2.0}
\cite{bib:Schneider2019}, using a Fourier-Chebyshev-Fourier discretisation for
the velocity field on a rectangular domain with periodic boundary condition in
stream- and spanwise directions and no-slip boundary conditions in the
wall-normal direction for the stream- and spanwise velocity component and a
nonzero vertical suction velocity for the wall-normal velocity component.  We
note that an adjustment in the suction velocity also results in a change in the
laminar profile through Eq.~\eqref{eq:laminar}.  

%That is, $\vec{U}(t)$ becomes
%time-dependent, and the application of $\eqref{eq:momentum}$ requires
%$\vec{U}(t)$ to obey the Navier-Stokes equations immediately after an
%adjustment of $V_s$.  
%\textbf{What does that mean? are all equations solved simultaneously, or 
%are $R$ and $u$ updated sequentially?}

The initial conditions are perturbations of an invariant solution $\vec{u}_0$
of Eqs.~\eqref{eq:momentum}-\eqref{eq:solenoidal}. Here we choose $\vec{u}_0$
to be a travelling wave with one unstable direction in a subspace that enforces
the symmetries of the invariant solution \cite{bib:Kreilos2013}.  All
simulations have been carried out %in the same domain as in
%Ref.~\cite{bib:Kreilos2013}, that is, 
with spanwise, wall-normal and streamwise
dimensions $L_x \times L_y \times L_z = 4\pi \times 8 \times 2\pi$ using $N_x
\times N_y \times N_z = 32 \times 65 \times 32$ grid points.  Further
simulation parameters are summarized in table \ref{tab:simulations}.

\begin{table}
  \begin{center}
  \begin{tabular}{cccccccc}
	  control variable & observable & $\Rey_0$ & $V_{s0}/U_0$ & $\mu$    & $\gamma$ & perturbation type    & $\|\delta \vec{u}_0\|_2/\|\vec{u}_0\|_2$ \\[3pt]
	  $V_s$            & $L_2$-norm & 1600      & 0.00058  & $-1.5 \times 10^7$ & 0.1      & random     & 0.01 \\
	  $V_s$            & $L_2$-norm & 1600      & 0.00058  & $-1.5 \times 10^3$ & 0.5      & mean flow  &  4.5 \\
	  $V_s$            & $C_f$      & 1600      & 0.00058  & $-10^{4}$ - $-10^5$  & 1.0      & random   &  1.5   
  \end{tabular}
  \caption{Simulation parameters and observables, with $V_{s}$ the suction velocity, $C_f$ the friction
	  coefficient, $\Rey_0 = U_0d/\nu$ the Reynolds number, $V_{s0}$ the suction velocity at the operating point,
	  $\mu$ and $\gamma$ the free parameters in Eq.~\eqref{eq:control} and $\delta \vec{u}_0$ 
	   a perturbation of the invariant solution $\vec{u}_0$.
          }
  \label{tab:simulations}
  \end{center}
\end{table}

\section{Results}
\label{sec:results}

\begin{figure}
%        {\includegraphics[width=0.31\columnwidth]{{figures/L2_vs_R-addUbase.eps}}}
%        {\includegraphics[width=0.31\columnwidth]{{figures/L2_vs_R.eps}}}
%        {\includegraphics[width=0.31\columnwidth]{{figures/Cf_vs_t.eps}}}
        {\includegraphics[width=0.31\columnwidth]{{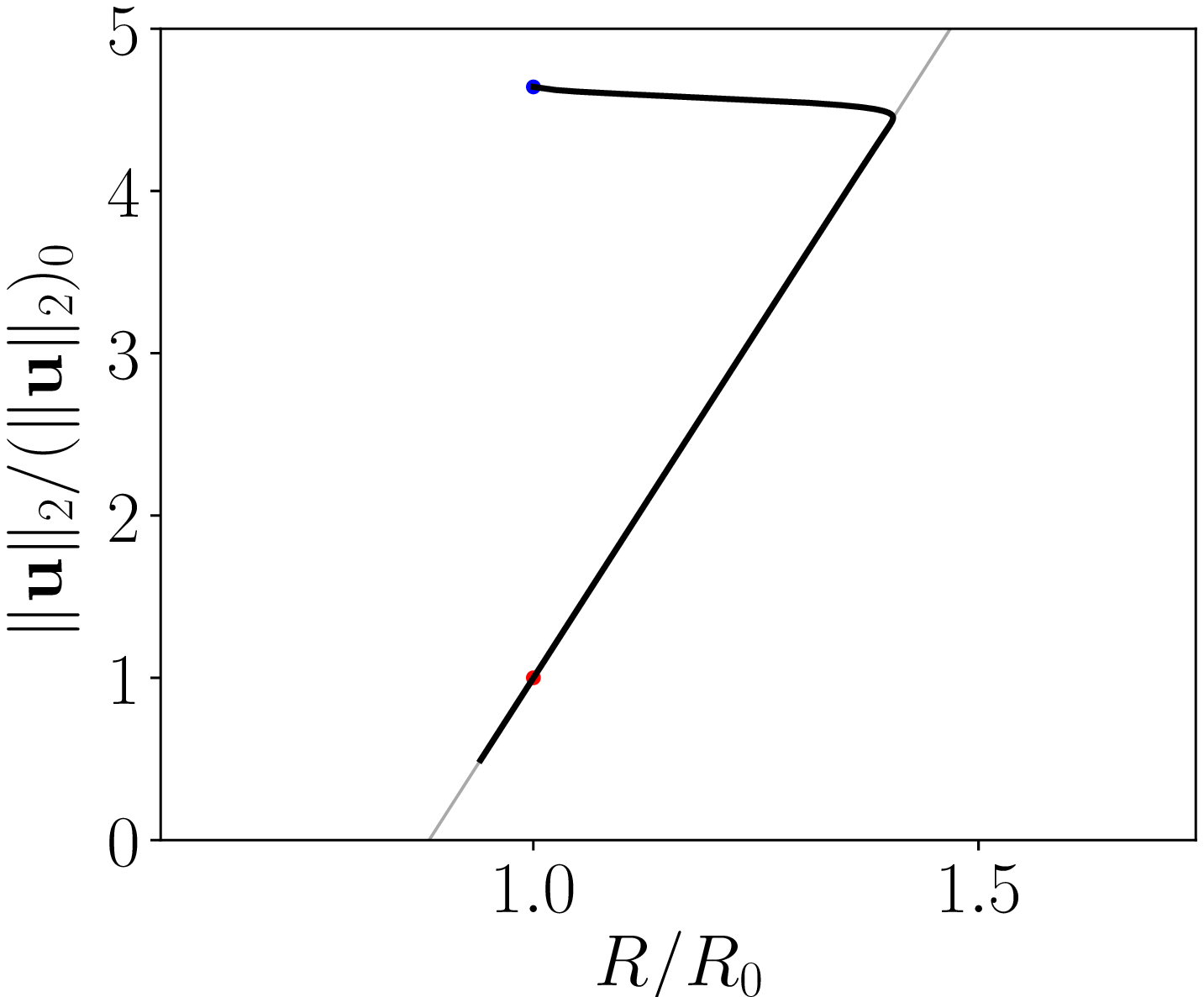}}}
        {\includegraphics[width=0.31\columnwidth]{{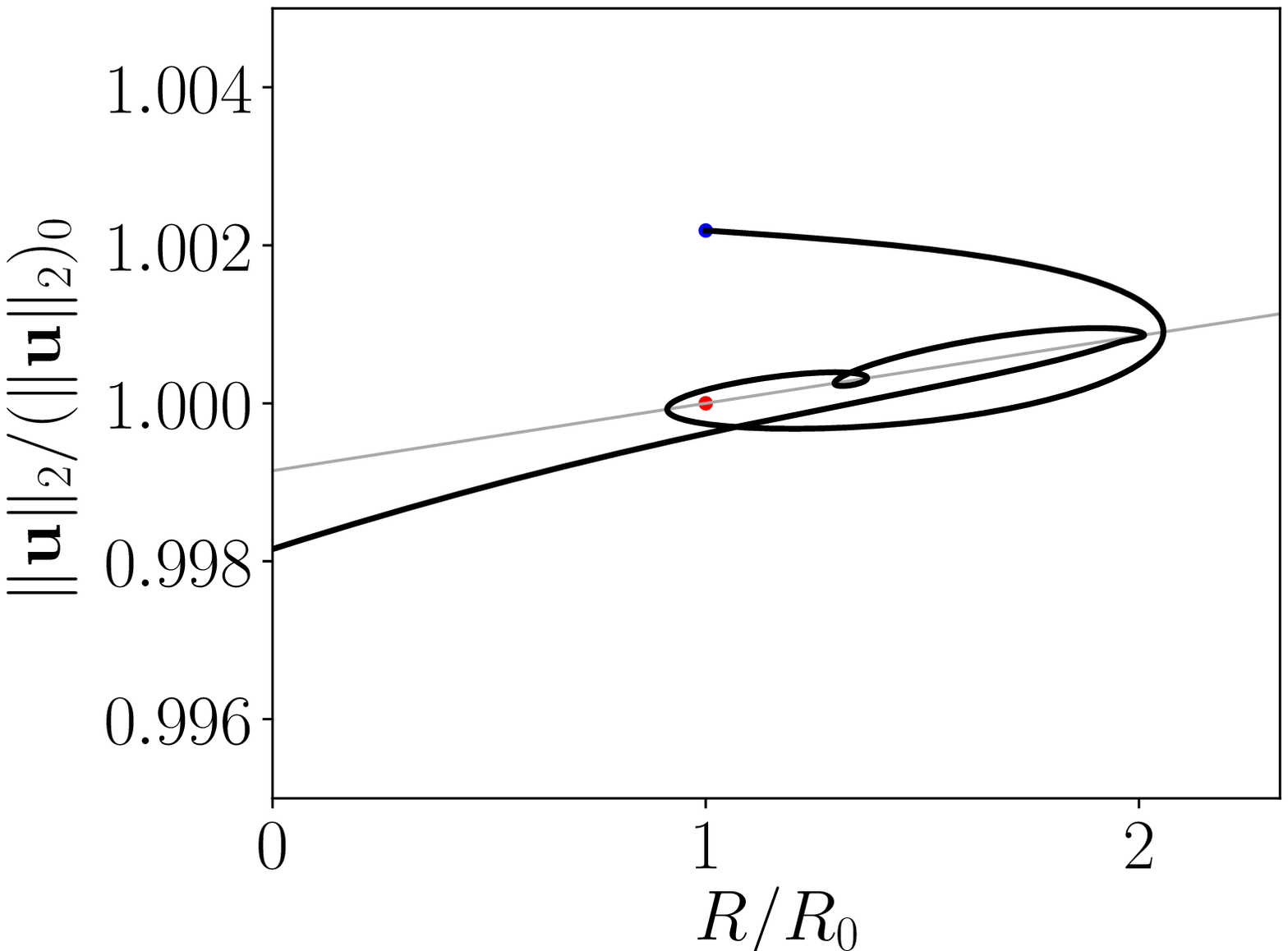}}}
        {\includegraphics[width=0.31\columnwidth]{{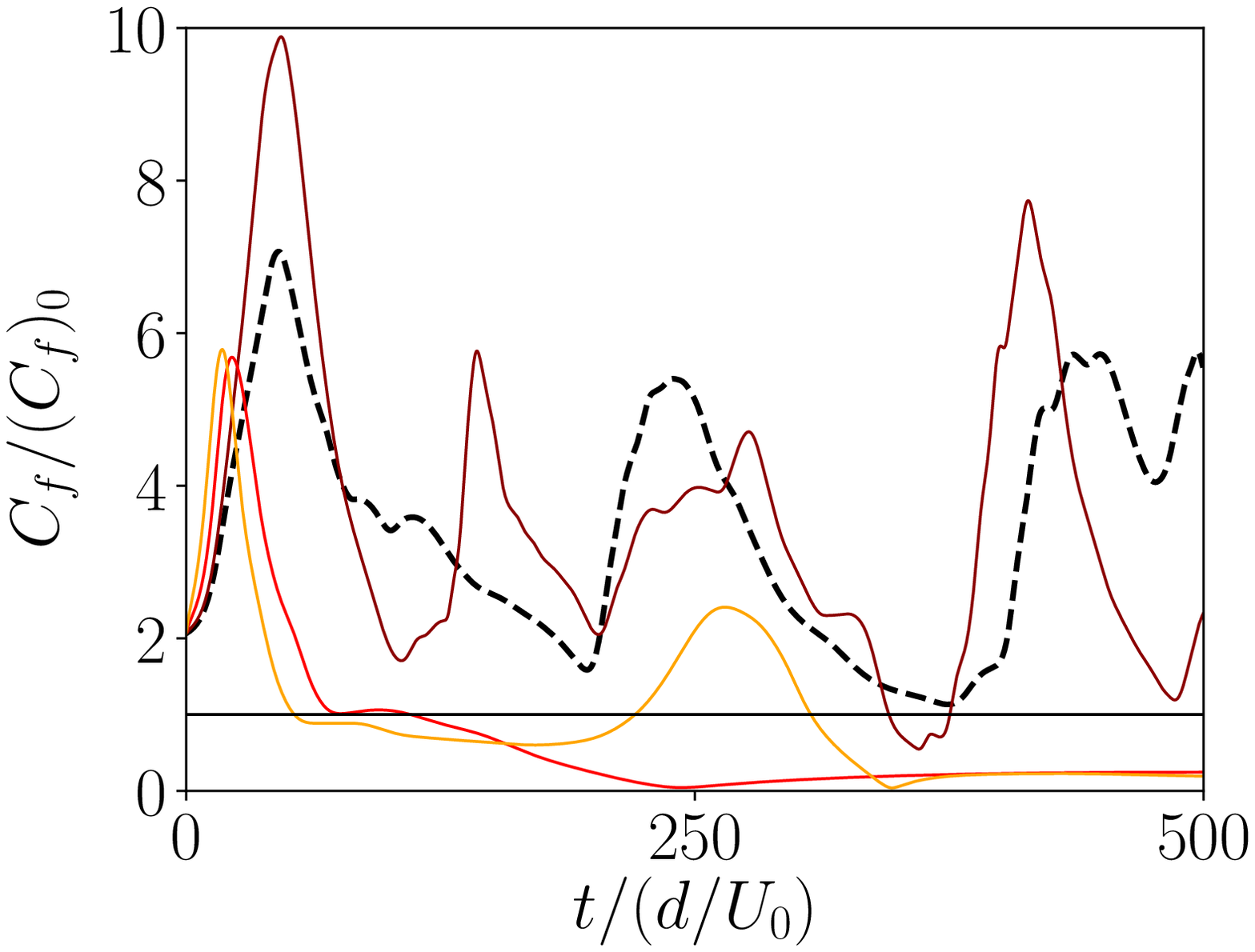}}}
\caption{
	Left and middle: Controlled dynamics of an invariant solution subject to perturbations of the mean flow (left) and 
	random perturbations (middle). The initial 
	data is indicated in blue and the operating point in red. 
	The control line with slope $1/\mu$ is shown in grey.
	Right: 
	Controlled dynamics for different values of $\mu$ (dark red: $\mu = -10^4$, red: $\mu = -5 \times 10^4$, 
	orange: $\mu = -10^5$) with a prescribed 
	value of the friction coefficient (solid black line).  
	The dashed line corresponds to the uncontrolled dynamics. 
}
\label{fig:phase-space-asbl}
\end{figure}

Applying a linear feedback control to an invariant solution with one unstable
direction can result in stabilisation thereof, if (i) the control overlaps with
the unstable direction, and (ii) if the stable directions are not destabilised.
Unstable invariant solutions have indeed been stabilised in pipe flow through
an instantaneous feedback control strategy where the Reynolds number is
adjusted in response to an observable connected with deviations from laminar
flow \cite{bib:Duguet2017}.  Here, we attempt to stabilise an unstable
travelling wave through the suction control procedure specified in
Eqs.~\eqref{eq:momentum}-\eqref{eq:control}.  
Since the feedback procedure is
based on the suction velocity, which is translationally invariant in all
directions, perturbations of the mean flow should be easiest to control.  
The left and middle panels of Fig.~\ref{fig:phase-space-asbl} show phase-space
representations of the controlled dynamics, starting from either a strong
perturbation of the mean flow (left panel) or a weak random perturbation
(middle panel).  One notes that the controlled dynamics approaches the operating
point in both examples. However, on longer times the systems slowly drifts away from
the target state, due to the adverse effects of the control on the stable directions
of the invariant solution \cite{bib:Linkmann2019}. A stable control hence requires
a feedback that is orthogonal to the stable directions.

As a second application of the feedback control, we use it to target a global constraint 
on the flow. Specifically, we pick a target value for the drag that is lower than the 
average drag $C_f = 5 \times 10^{-3}$  at a Reynolds number of $\Rey_0 = 1600$. % \textbf{give values},
%
%certain areas of phase space can be avoided by choosing the target value of the
%observable accordingly, thereby avoiding high-drag states, for instance.  
The right most panel in Fig.~\ref{fig:phase-space-asbl} shows time series of the
controlled for three different values of $\mu$ 
and the free dynamics, where the target value of the shear stress was set to $(C_f)_0 = 1.25 \times 10^{-3}$. 
The uncontrolled dynamics shows large fluctuations. The controlled dynamics 
shows much smaller fluctuations if $\mu$ is large enough, 
approaches the target value and stays its vicinity for up to $200$ time
units. 
%asymptotically approaches the target value.

%This time, there is no evidence for a long time destabilization or run away from
%the target value \textbf{Stimmt das so? Sonst: up to times ....}.

%\begin{acknowledgement}
%  An acknowledgement may be placed at the end of the article.
%\end{acknowledgement}

\vspace{\baselineskip}
%% The style of the following references should be used in all documents.


\begin{thebibliography}{1}

\bibitem{bib:Schneider2019}% 
J.~F. Gibson, F. Reetz, S. Azimi, A. Ferraro, T. Kreilos, H. Schrobsdorff, M. Farano, A.~F. Yesil, S.~S. Schütz, M. Culpo \& T.~M. Schneider, 
"Channelflow 2.0", manuscript in preparation (2019), see {\tt channelflow.ch}

\bibitem{bib:Kreilos2013}% 
T. Kreilos, G. Veble, T.~M. Schneider \& B. Eckhardt,
%Edge states for the turbulence transition in the asymptotic suction boundary layer
J. Fluid Mech.  {\bf 726},  100--122 (2013)  

\bibitem{bib:Duguet2017}% 
A.~P. Willis, Y. Duguet, O. Omel'chenko, \& M. Wolfrum, 
%{Surfing the edge: using feedback control to find nonlinear solutions}  
J. Fluid Mech.  {\bf 831},  579--591 (2017)  

\bibitem{bib:Linkmann2019}% 
M. Linkmann, F. Knierim, S. Zammert \& B. Eckhardt,
manuscript in preparation (2019)
\end{thebibliography}
\end{document}